\documentclass{pazh_engl}

\usepackage{flushrt}
\usepackage{graphicx}

\begin{document}

\sloppypar

\title{\bf A Hard X-ray survey of the Galactic Center with INTEGRAL/IBIS.
Catalog of sources}

\author{\copyright 2004 C,. M.Revnivtsev\inst{1,2}, R.Sunyaev\inst{1,2}, 
D.Varshalovich\inst{3}, V.Zheleznyakov\inst{4}, A.Cherepashchuk\inst{5}, 
A.Lutovinov\inst{1},E.Churazov\inst{1,2}, S.Grebenev\inst{1}, 
M.Gilfanov\inst{1,2}
}

\institute{Space Research Institute, Moscow, Russia
\and
Max-Plank-Institut fuer Astrophysik, Garching, Germany
\and
Ioffe Physico-Technical Institute, St.-Peterburg, Russia
\and
Institute of Applied Physics, N.Novgorod, Russia
\and
Sternberg Astronomical Institute, Moscow, Russia
}
\authorrunning{REVNIVTSEV ET AL.}
\titlerunning{INTEGRAL/IBIS GALACTIC CENTER HARD X-RAY SURVEY}

\date{Jan.19 2004}

\abstract{
During the period Aug.23-Sept.24 2003, the INTEGRAL observatory
performed an ultra deep survey of the Galactic Center region with a record 
sensitivity at energies higher than 20 keV. We have analized images of
the Galactic Center region obtained with the ISGRI detector of the
IBIS telescope (15-200 keV) and present here a catalog of detected sources. 
In total, 60 sources with a flux higher than 1.5 mCrab have been
detected. 44 of them were earlier identified as Galactic binary systems, 3 are
extragalactic objects. 2 new sources are discovered. 
}
\maketitle

\section*{INTRODUCTION}

Bright Galactic X-ray sources, most of which are binary systems 
with an accreting black hole or a neutron star, strongly concentrate
towards the Galactic plane and especially towards the Galactic Center.
Therefore, imaging instruments are very important in studying the
X-ray emission from these objects. For some time in the past, X-ray
images of the sky were obtained from scanning observations with detectors
having collimator-limited fields of view (e.g. UHURU, Forman et
al. 1978, HEAO1, Wood et al. 1984). The advent of X-ray telescopes
using the grazing reflection technique (e.g. EINSTEIN, ROSAT, CHANDRA,
XMM) dramatically increased the resolution of X-ray images. However,
neither of these telescopes can produce images at energies higher than
$\sim$10 keV. 

Focusing hard X-ray photons with energies above 30 keV is very
difficult. Therefore, telescopes used for image reconstruction in this energy
range employ another method  -- coded mask aperture (see e.g. Fenimore,
Cannon 1978; Skinner et al. 1987a). A number of telescopes built on
this principle have obtained very interesting results, in
particular for the Galactic Center region, e.g. Spacelab/XRT (Skinner
et al. 1987b), MIR/KVANT/TTM(Sunyaev et al. 1990),
GRANAT/ART-P(Pavlinsky et al. 1992, 1994), GRANAT/SIGMA(Cordier et
al. 1991, Sunyaev et al. 1991), BeppoSAX/WFC (Ubertini et al. 1999) 

The last survey of the Galactic Center region at energies above 40 keV
was completed by the SIGMA telescope aboard the GRANAT observatory more than
5 years ago. More than 5 million sec were spent on observations of
this region, which made it possible to achieve a sensitivity of
$\sim$5 mCrab in the energy band 40-150 keV (e.g. Churazov et al. 1994). Now
with the IBIS telescope of the INTEGRAL observatory we have the possibility
to carry out a new hard X-ray  survey of the Galactic Center at a much
higher sensitivity in the energy band 20-150 keV.

In this article we present an analysis of data obtained with
INTEGRAL/IBIS during an ultra deep survey of the Galactic Center
region that was performed by the INTEGRAL observatory in August-September
2003. The total exposure time of the utilized observations is approximately 2
million sec. We analize images of the region and present a catalog
of detected sources. More detailed analysis of the X-ray emission of
the detected sources will be presented in future papers.

\section*{INSTRUMENT AND DATA ANALYSIS}

The INTEGRAL observatory (Winkler et al. 2003) was launched by a Russian
rocket-launcher PROTON Oct.17 2002 (Eismont et al. 2003).
There are four telescopes on board the observatory. In the present
work, we will only use data from the detector ISGRI of the telescope
IBIS (Ubertini et al. 2003). This instrument provides the best combination
of field of view, sensitivity and angular resolution for our study. 
The lower detector of the telescope IBIS - PICsIT has an effective
energy range starting at $\sim$170 keV and is therefore 
less sensitive to typical X-ray sources.

The coded-mask telescope IBIS has a field of view of $29^\circ\times29^\circ$ 
(the fully coded field of view is $9^\circ\times9^\circ$) and makes it
possible to construct images with an angular resolution of
$\sim12^\prime$. The detector ISGRI mounted on this telescope consists of 16384
independent pixels of the CdTe semiconductor. A detailed description of the
detector can be found in (Lebrun et al. 2003).

All the available data were processed in a uniform way. For each Science
Window, the event energies were calculated following the prescription
implemented in OSA 3.0 (Goldwurm et al. 2003) using the gain and rise time
correction tables versions 9 and 7 respectively. Events, accumulated
in raw detector coordinates, were screened for possible hot or dead
pixels, which resulted in a rejection of a few percent of the pixels. The raw
detector images in the 18-60 keV band (which provides the best
sensitivity for typical X-ray sources) were then rebinned into a new
grid having a pixel size equal to $1/3$ of the mask pixel size. This
causes inevitable (although modest) loss of spatial resolution, but
has an advantage of straightforward application of standard coded
masks reconstruction algorithms such as DLD deconvolution
(e.g. Fenimor, Cannon 1981, Skinner et al. 1987a). The "balanced" DLD
deconvolution was then applied to detector images. A balance matrix
was calculated to compensate the spatially nonuniform background across
the detector. The background used for balance matrix calculations was
accumulated over a large set of observations so the contribution of
individual sources to the background is completely smeared. An iterative
source removal procedure was used to eliminate ghosts of sources due
to incomplete coding. The resulting DLD images were then coadded to
form a combined map. The map was finally convolved with a Gaussian
having a width approximatelly matching the width of the real PSF, which
is somewhat broader then the ideal DLD PSF due to the detector
rebinning and co-adding multiple images with different
orientations. 

The same procedure was applied to a big set of Crab Nebula
observations, thus allowing direct check of the applied method and
comparison of detected fluxes with standard "Crab" units.
We have concluded from this analysis that the current software allows
recovering absolute source fluxes with $\sim$10 \% uncertainty. The
accuracy of bright source localization is approximately 0.4$^\prime$ (1$\sigma$
confidence contour). For weaker sources this error radius can increase
to 2-3$^\prime$.

The average image of the Galactic Center region was searched for sources.
A peak on the image with an intensity higher than a specified limit
was regarded as indicating the presence of a point source. We adopted
a 6.5$\sigma$ detection limit, which is slightly higher than allowed
by the statistics. This allows us to avoid dealing with systematic
uncertainties that arise in analyzing images of such crowded regions
as the Galactic Center with the present software. The sensitivity map
derived with the adopted detection limit is presented in Fig.1.

\begin{figure}[htb]
\includegraphics[width=\columnwidth]{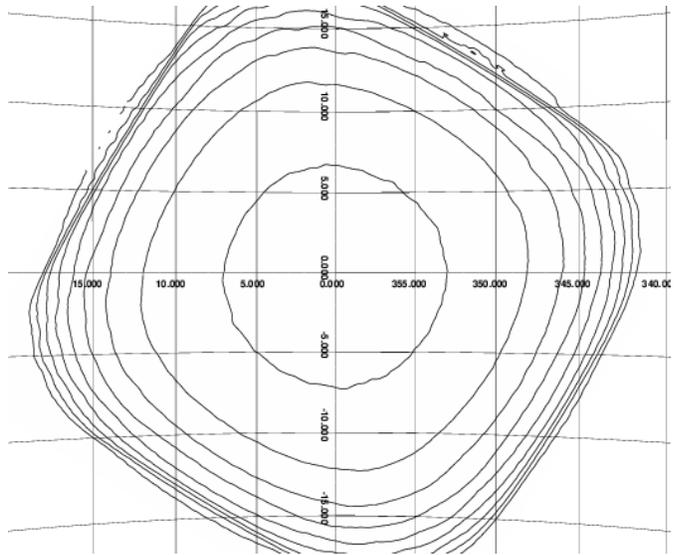}
\caption{Sensitivity map corresponding to the analized observations.
Contours correspond to source detection limits of 1, 1.5, 2.0, 2.5 etc
mCrab}
\end{figure}

\section*{RESULTS}
A list of detected sources, their coordinates and fluxes in the 18-60 keV
energy band are presented in Table 1. A flux of 1 mCrab in this energy
band corresponds to $\sim1.4\times 10^{-11}$ ergs/s/cm$^2$ for sources
with a Crab-like spectrum. The accuracy of source localization
is approximately 2--3$^\prime$ (90\% confidence contour).

\begin{figure*}[t]
\includegraphics[width=\textwidth]{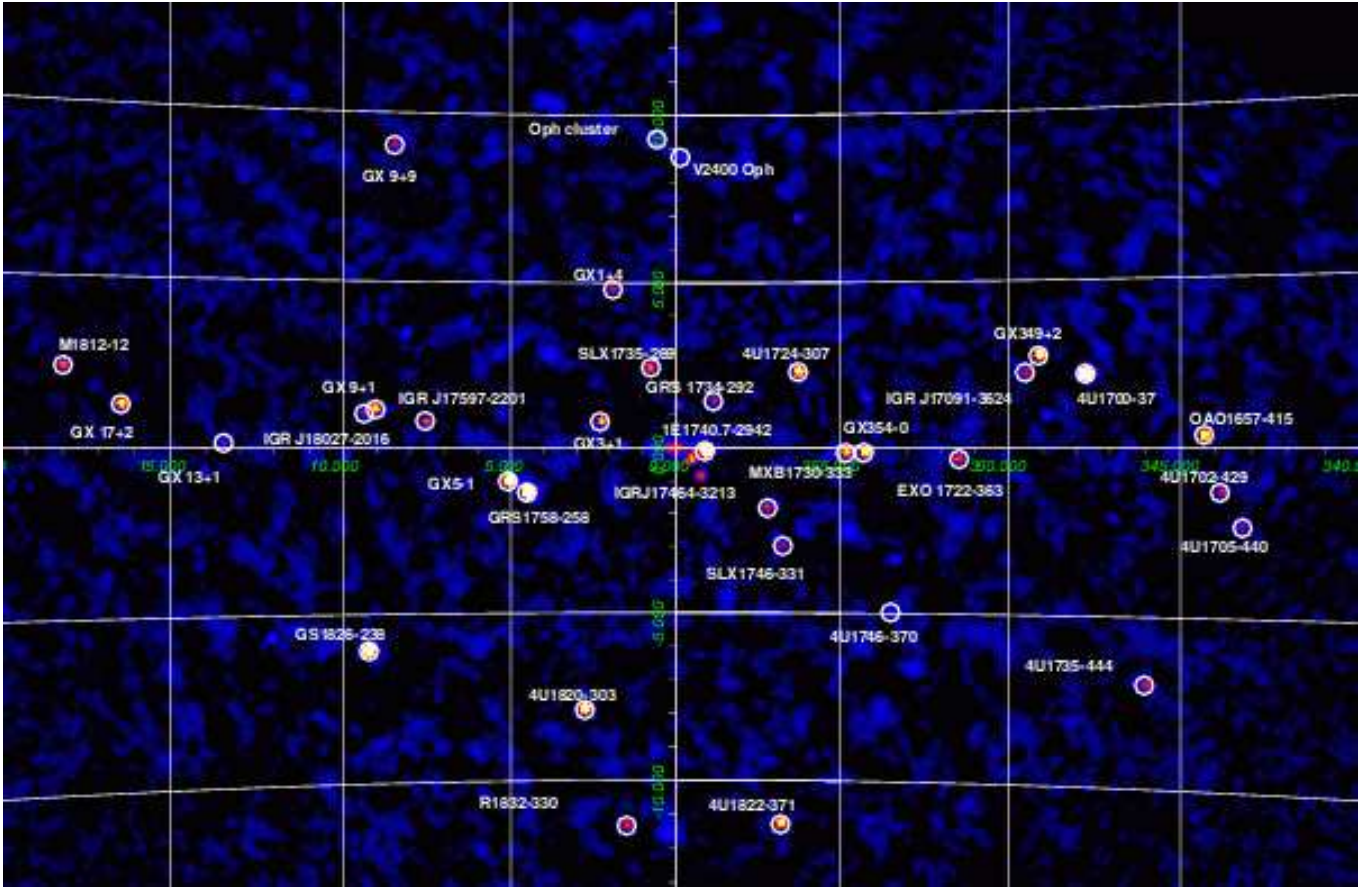}
\caption{Averaged map of the Galactic Center region in the energy band 
18-60 keV. The total exposure time is $\sim$ 2 million sec}
\end{figure*}

An image of the Galactic Center region (size $\sim 35^\circ \times 25^\circ$)
is presented in Fig.1. In order to avoid too many subscripts on the image,
we labeled only part of the detected sources. Zoomed images of the
Galactic Center ($\sim 7^\circ \times 5^\circ$ size) obtained from 
INTEGRAL/IBIS/ISGRI (18-60 keV) and MIR/KVANT/TTM(2-20 keV) data
are presented in Fig. 3. The location of the dynamical center of the Galaxy
(the supermassive black hole Sgr A$^*$) is shown by a cross. Because
of the limited angular resolution of the IBIS and TTM telescopes we cannot
determine the flux of the Sgr A$^*$ itself (for more details see
Baganoff et al. 2003, Belanger et al. 2003).

By analyzing the averaged 18-60 keV image of the Galactic Center
region with a typical sensitivity of 1-2 mCrab we have detected
60 sources. The nature of 50 of them is known, most of them (37 of 50)
being low mass X-ray binaries (LMXB). A significant fraction of these
binary systems are bursters (i.e. binary systems with neutron stars,
episodically demonstrating unstable thermonuclear burning -- bursts --
on the surface of the neutron star). However, bright neutron star
systems that do not demonstrate such bursting behavior (GX5-1, GX9+1,
GX9+9, GX13+1, GX340+0) are also present in the field.

Also detected are an anomalous X-ray pulsar -- 1RXS J170849.0-40091
and 4 binary X-ray pulsars -- OAO1657-415, AXJ1749.2-2725, EXO1722-363
and GX1+4. Except for the last one, these sources are high mass X-ray binary
systems (HMXB). 

Two detected sources (V1223 Sgr and V2400 Oph) are cataclysmic variables (CV),
one (Oph cluster=4U1708-23) is a cluster of galaxies, one (GRS
1734-292) is an active galactic nucleus, one (PKS 1830-211) is a lensed blazar 
($z\approx 2.5$), and one (SGR 1806-20) is a soft gamma repeater.

Among the known binary systems, only four (1E1740.7-2942, GRS
1758-258, SLX 1746-331 and IGR J17464-3213/H1743-31) harbor black hole
candidates (BH).

Five sources from Table 1 were discovered by INTEGRAL, three of them
during these observations (IGR J17544-2619, IGR J18027-2016, IGR
J17475-2822).  
\footnote{ IGR J17475-2822 is located only
 0.7$^\circ$ from the Galactic Center and can be a superposition of 
weaker sources. Observations with higher angular resolution are needed}
IGR J17544-2619 is a transient source, which was detectable only
 during part of our observations (Sunyaev et al. 2003, Grebenev et
 al. 2003), it cannot be detected on the averaged map of the region.

Therefore, only 10 of the 60 sources detected in the Galactic Center
region are of unknown origin. Four of them have positions coincident
with sources previously discovered by the ROSAT observatory. 

\bigskip

Work was partly supported by a grant of Minpromnauka (grant of president of 
Russian Federation NSH-2083.2003.2)  and the programm of the Russian
Academy of Sciences 'Non-stable phenomena in astronomy''. Authors thank
the International Science Data Center of INTEGRAL (Versoix, Swiss) and
Russian Science Data Center of INTEGRAL (Moscow, Russia). 
The work is based on observations with INTEGRAL, an ESA project with
instruments and science data centre funded by ESA member states
(especially the PI countries: Denmark, France, Germany, Italy,
Switzerland, Spain), Czech Republic and Poland, and with the
participation of Russia and the USA.

\section*{REFERENCES}

Baganoff F., Maeda Y., Morris M. et al.// Astroph.J. {\bf 591}, 891 (2003)

Belanger G., Goldwurm A., Goldoni P. et al.// astro-ph/0311147  (2003)

Churazov E., Gilfanov M., Sunyaev R. et al. // Astroph.J.Suppl.{\bf 92}, 381 
(1994)

Cordier B., Goldwurm A., Laurent P.  et al.// Adv.Sp.Res. {\bf 11}, 169 (1991)

Eismont N.A., Ditrikh A.V.,  Janin G. et al.// Astron. Astroph. {\bf 411}, 37 
(2003)

Fenimore E. E., Cannon T. M.// Appl.Opt. {\bf 17}, 337 (1978)

Fenimore E. E., Cannon T. M.// Appl.Opt. {\bf 20}, 1858 (1981)

Forman W., Jones C., Cominsky L. et al. // Astroph.J.Suppl. {\bf 38}, 357 
(1978)

Goldwurm A., David P., Foschini L. et al.// Astron. Astroph. {\bf 411}, 223 
(2003)

Grebenev S., Lutovinov A., Sunyaev R.// Astronomer's Telegram 192 (2003)

Lebrun F., Leray J. P., Lavocat P. et al.// Astron. Astroph. {\bf 411}, 141 
(2003)

Pavlinsky M., Grebenev S., Sunyaev R.// Sv.AL {\bf 18}, 291 (1992)

Pavlinsky M., Grebenev S., Sunyaev R. // Astroph.J. {\bf425}, 110 (1994)

Skinner G. K., Ponman T. J., Hammersley A. P., Eyles, C. J.// Astroph. and 
Space Science {\bf 136}, 337 (1987a)

Skinner G. K., Willmore, A. P., Eyles C. J., Bertram D., Church M. J.// Nature
 {\bf 330}, 544 (1987b)

Sunyaev R. A., Churazov E., Efremov V., Gilfanov M., Grebenev S.// Adv.Sp.Res. 
{\bf 10}, 41 (1990)

Sunyaev R., Churazov E., Gilfanov M. et al.// Astron. Astroph. {\bf 247}, 29 
(1991)

Sunyaev R., Grebenev S., Lutovinov A. et al.// Astronomer's Telegram 190 (2003)

Ubertini P., Bazzano A., Cocchi M. et al.// Astroph.Lett.Comm. {\bf 38}, 301 
(1999)

Ubertini P., Lebrun F., Di Cocco G. et al.// Astron. Astroph. {\bf 411}, 131 
(2003)

Winkler C., Courvoisier T. J.-L., Di Cocco G. et al.// Astron. Astroph. {\bf 
411}, 1 (2003)

Wood K.S., Meekins J.F., Yentis D.J. et al.// Astroph.J.Suppl. {\bf 56}, 507 
(1984)

\begin{table*}
\fontsize{9pt}{10pt}\selectfont
\caption{List of sources detected on the averaged map of the Galactic
Center region, obtained from the ultra deep survey performed by INTEGRAL
in Aug.-Sep.2003.} 

\begin{tabular}{l|c|c|c|c|l|l}
\hline
&$\alpha$(2000)&$\delta$(2000)&Flux (18-60 keV), mCrab $^a$ & $\sigma$ & 
Identification&Class\\
\hline
1 & 255.978 & -37.841 & $ 187.5 \pm   0.3 $ & 662.5 & 4U1700-377        &HMXB, 
NS   \\
2 & 265.977 & -29.746 & $  74.8 \pm   0.2 $ & 491.8 & 1E1740.7-2942      
&LMXB, BH  \\
3 & 270.308 & -25.748 & $  78.0 \pm   0.2 $ & 478.2 & GRS 1758-258       
&LMXB, BH  \\
4 & 277.370 & -23.805 & $  93.6 \pm   0.2 $ & 392.4 & GS1826-238         
&LMXB, NS  \\
5 & 270.289 & -25.080 & $  42.7 \pm   0.2 $ & 259.0 & GX5-1              
&LMXB, NS  \\
6 & 275.917 & -30.367 & $  32.3 \pm   0.2 $ & 169.4 & 4U1820-303         
&LMXB, NS  \\
7 & 256.424 & -36.416 & $  38.9 \pm   0.2 $ & 160.5 & GX349+2            
&LMXB, NS  \\
8 & 261.887 & -30.804 & $  24.5 \pm   0.2 $ & 151.6 & 4U1724-307 (Ter 2) 
&LMXB, NS  \\
9 & 262.981 & -33.831 & $  22.9 \pm   0.2 $ & 135.8 & GX354-0            
&LMXB, NS  \\
10 & 255.187 & -41.658 & $  59.7 \pm   0.5 $ & 126.6 & OAO1657-415       
&HMXB, NS   \\
11 & 276.438 & -37.113 & $  29.0 \pm   0.2 $ & 117.6 & 4U1822-371        
&LMXB, NS   \\
12 & 263.354 & -33.381 & $  14.6 \pm   0.2 $ &  88.4 & RAPID BURSTER     
&LMXB, NS   \\
13 & 266.991 & -26.566 & $  12.6 \pm   0.2 $ &  81.7 & GX3+1             
&LMXB, NS   \\
14 & 270.387 & -20.523 & $  15.8 \pm   0.2 $ &  77.2 & GX9+1             
&LMXB, NS   \\
15 & 266.519 & -29.514 & $  10.4 \pm   0.2 $ &  69.3 & A1742-294         
&LMXB, NS\\
16 & 264.560 & -26.993 & $  10.0 \pm   0.2 $ &  64.4 & SLX 1735-269      
&LMXB, NS   \\
17 & 274.014 & -14.037 & $  37.3 \pm   0.6 $ &  63.1 & GX17+2            
&LMXB, NS   \\
18 & 266.834 & -30.010 & $   7.6 \pm   0.2 $ &  50.0 & SLX1744-299/300   
&LMXB, NS   \\
19 & 261.286 & -36.280 & $   9.5 \pm   0.2 $ &  48.4 & EXO 1722-363      
&HMXB?, NS   \\
20 & 269.958 & -22.019 & $   8.6 \pm   0.2 $ &  46.6 & IGR/XTE J17597-2201   
&LMXB, NS   \\
21 & 266.543 & -32.235 & $   7.2 \pm   0.2 $ &  45.8 & IGR 
J17464-3213/H1743-322  &XB,BH?    \\
22 & 264.729 & -44.438 & $  24.5 \pm   0.6 $ &  43.9 & 4U1735-444        
&LMXB,NS  \\
23 & 278.921 & -32.995 & $  10.9 \pm   0.2 $ &  43.7 & R1832-330 (NGC 6652)  
&LMXB, NS   \\
24 & 256.540 & -43.043 & $  24.5 \pm   0.6 $ &  43.6 & 4U1702-429        
&LMXB, NS   \\
25 & 266.491 & -28.923 & $   6.2 \pm   0.2 $ &  41.0 & 1E1742.8-2853      
&LMXB   \\
26 & 257.290 & -36.394 & $   9.1 \pm   0.2 $ &  39.4 & IGR J17091-3624   &?   
\\
27 & 266.404 & -29.018 & $  57.4 \pm   0.2 $ &  37.9 & AX J1745.6-2901   & 
LMXB\\
28 & 266.580 & -28.735 & $   5.6 \pm   0.2 $ &  37.3 & 1E1743.1-2843 &LMXB \\
29 & 264.364 & -29.137 & $   5.3 \pm   0.2 $ &  35.0 & GRS1734-292       &AGN  
\\
30 & 263.003 & -24.747 & $   5.7 \pm   0.2 $ &  34.3 & GX1+4             
&LMXB, NS   \\
31 & 267.466 & -33.192 & $   5.4 \pm   0.2 $ &  33.8 & SLX 1746-331        
&LMXB?  \\
32 & 262.905 & -16.991 & $   9.9 \pm   0.3 $ &  31.7 & GX9+9             
&LMXB, NS   \\
33 & 257.210 & -44.089 & $  17.0 \pm   0.7 $ &  25.9 & 4U1705-440         
&LMXB, NS  \\
34 & 270.690 & -20.272 & $   4.8 \pm   0.2 $ &  23.0 & IGR J18027-2016(new)&?  
\\
35 & 273.630 & -17.149 & $   7.6 \pm   0.3 $ &  22.8 & GX13+1             
&LMXB, NS  \\
36 & 265.152 & -28.303 & $   3.4 \pm   0.2 $ &  22.4 & SLX 1737-282       
&LMXB  \\
37 & 273.780 & -12.103 & $  23.8 \pm   1.2 $ &  20.6 & M1812-12           
&LMXB, NS\\
38 & 258.140 & -23.334 & $   4.2 \pm   0.2 $ &  20.1 & Oph cluster        
&Cluster \\
39 & 267.511 & -37.061 & $   3.5 \pm   0.2 $ &  18.1 & 4U1746-370 (NGC 6441) 
&LMXB, NS \\
40 & 257.528 & -28.162 & $   3.5 \pm   0.2 $ &  18.1 & XTEJ1710-281         
&LMXB  \\
41 & 258.159 & -24.232 & $   3.2 \pm   0.2 $ &  15.8 & V2400 Oph           &CV 
\\
42 & 266.261 & -28.902 & $   2.3 \pm   0.2 $ &  15.4 & GRS 1741.9-2853     
&LMXB, NS\\
43 & 266.868 & -28.370 & $   2.2 \pm   0.2 $ &  14.8 & IGR 
J17475-2822(new*)&?\\
44 & 261.317 & -32.976 & $   2.5 \pm   0.2 $ &  14.8 & 1RXS J172525.5-325717  
&?\\
45 & 266.205 & -29.352 & $   2.1 \pm   0.2 $ &  14.1 & KS1741-293             
&LMXB, NS\\
46 & 274.652 & -17.040 & $   4.9 \pm   0.4 $ &  13.6 & SAX J1818.6-1703     
&LMXB?\\
47 & 264.802 & -30.329 & $   2.0 \pm   0.2 $ &  13.1 & IGR/XTE J17391-3021     
 &HMXB?\\
48 & 272.148 & -20.446 & $   2.8 \pm   0.2 $ &  12.8 & SGR 1806-20     &SGR\\ 
49 & 267.335 & -27.511 & $   1.9 \pm   0.2 $ &  12.7 & AX J1749.2-2725 &HMXB, 
NS\\
50 & 267.828 & -20.174 & $   2.5 \pm   0.2 $ &  12.2 & 1RXS J175113.3-201214  
&?\\
51 & 257.294 & -32.302 & $   2.4 \pm   0.2 $ &  12.1 & 4U1705-32            
&?\\
52 & 258.073 & -37.652 & $   2.9 \pm   0.2 $ &  12.0 & SAX J1712.6-3739        
&LMXB, NS\\
53 & 278.379 & -21.084 & $   3.3 \pm   0.3 $ &  10.8 & PKS 1830-211           
&Blazar\\
54 & 260.027 & -31.288 & $   1.6 \pm   0.2 $ &   9.2 & 1RXS J172006.1-311702  
&?\\
55 & 265.745 & -36.345 & $   1.7 \pm   0.2 $ &   9.0 & XTE J1743-363  &?\\ 
56 & 251.452 & -45.609 & $  17.5 \pm   1.9 $ &   9.0 & GX340+0              
&LMXB, NS\\
57 & 283.722 & -31.133 & $   4.1 \pm   0.5 $ &   8.8 & V1223 Sgr            
&CV\\
58 & 266.590 & -21.543 & $   1.3 \pm   0.2 $ &   7.0 & 1RXS J174607.8-213333  
&?\\
59 & 257.245 & -40.111 & $   2.2 \pm   0.3 $ &   6.5 & 1RXS J170849.0-40091 
&AXP, NS\\
60$^b$ & 268.585 & -26.335 & $  83.1 \pm   2.0 $ &  30.0 & IGR J17544-2619 &?\\
\end{tabular}

\begin{list}{}
\item $^a$ -- here are quoted only statistical errors. Systematic
uncertainties ($\sim$10\%) are not included

\item $^b$ -- this source was detected only during part of the
observations. We quote maximum detection significance and maximum source flux

\end{list}

\end{table*}

\begin{figure*}
\begin{center}
  \includegraphics[height=0.45\textheight]{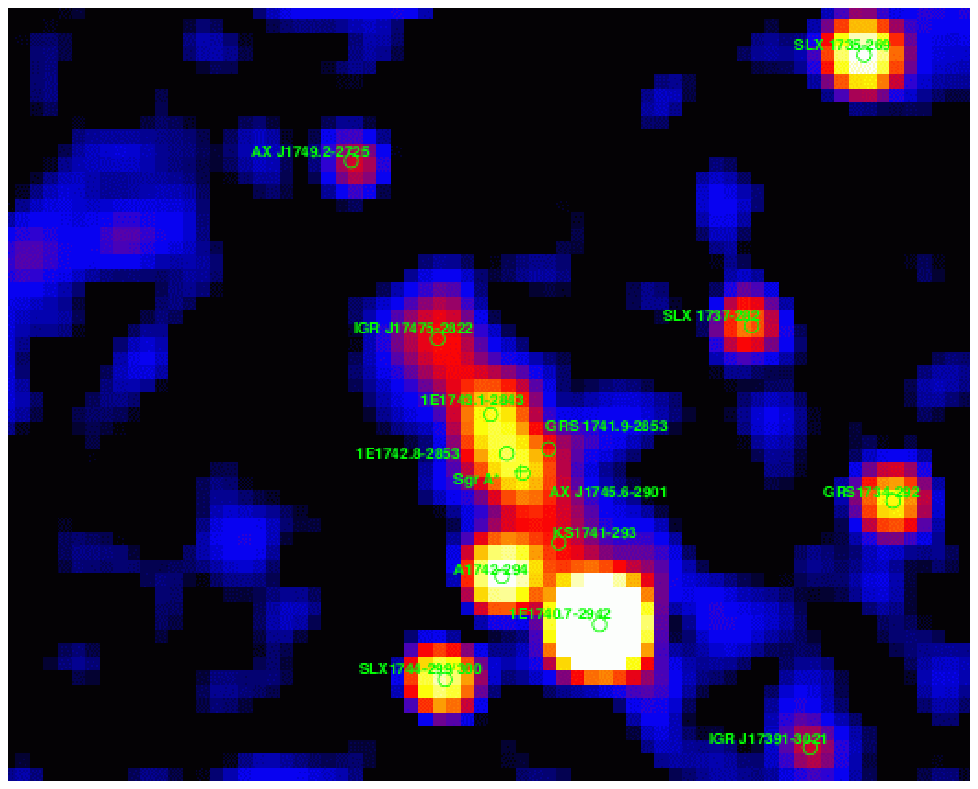}

  \includegraphics[height=0.45\textheight]{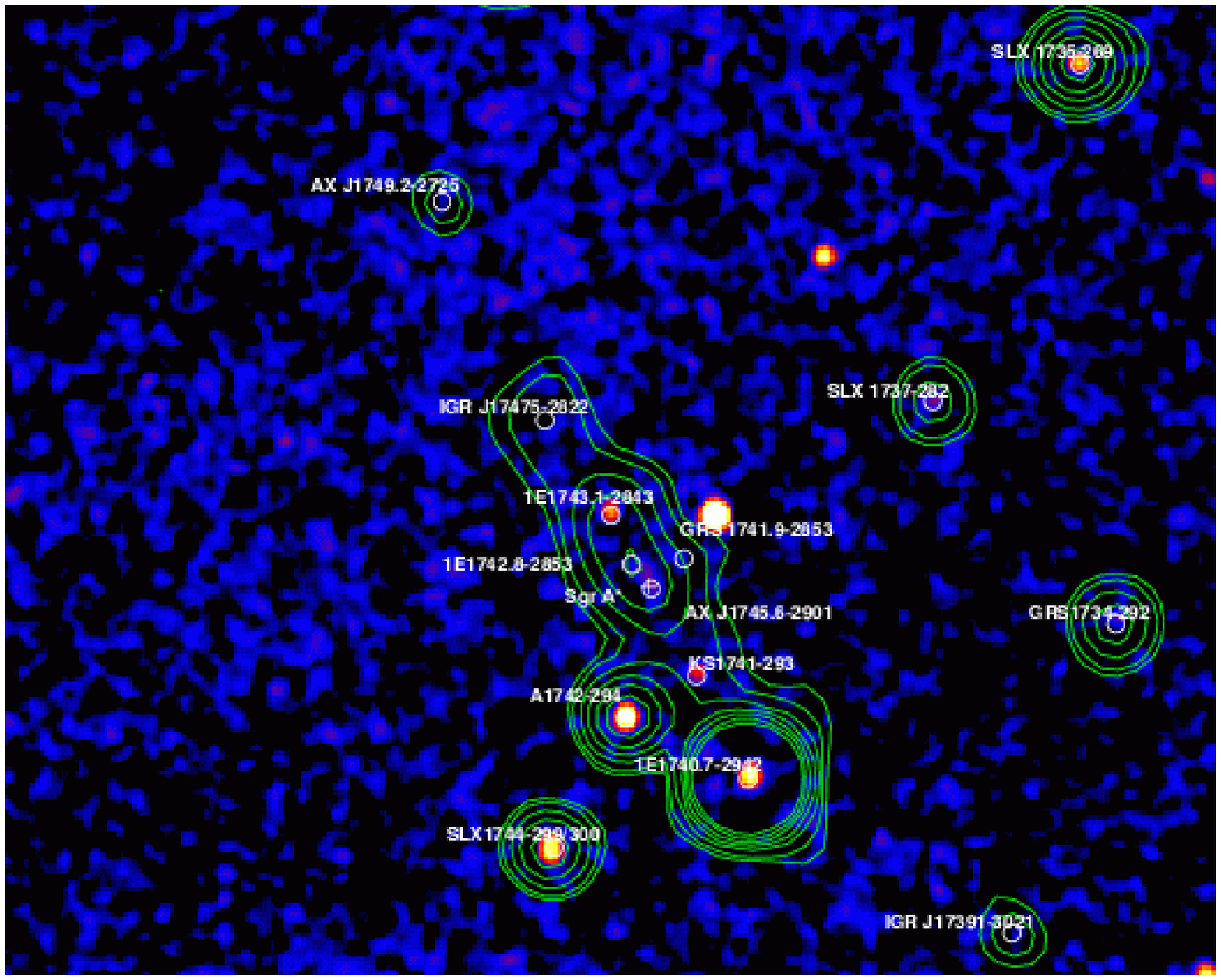}
\end{center}
\caption{Maps of the Galactic Center region ($\sim 7^\circ \times 5^\circ$).
The upper map collects data of the INTEGRAL/IBIS telscope (18-60 keV),
the lower map is based on data of MIR/KVANT/TTM (2-20 keV). Contours
on the lower image represent the brightness profile of the INTEGRAL/IBIS map}
\end{figure*}

\end{document}